\documentclass[aps,showpacs,manuscript,12pt]{revtex4}
\usepackage{amssymb}
\usepackage{amsmath}
\usepackage{graphicx}

\begin{document}

\title{\textbf{On the validity of the LAD and LL classical
radiation-reaction equations$^{\S }$ }}
\author{M. Dorigo$^{a}$, M. Tessarotto$^{b,c}$, P. Nicolini$^{b,c}$ and A.
Beklemishev$^{d}$}
\affiliation{\ $^{a}$Department of Physics, University of Trieste, Italy, $^{b}$%
Department of Mathematics and Informatics, University of Trieste, Italy, $%
^{c}$Consortium of Magneto-fluid-dynamics, University of Trieste, Italy, $%
^{d}$Budker Institute of Nuclear Physics, Novosibirsk, Russia}

\begin{abstract}
The search of the correct equation of motion for a classical charged
particle under the action of its electromagnetic (EM) self-field, the
so-called \textit{radiation-reaction equation of motion}, remains elusive to
date. In this paper we intend to point out why this is so. The discussion is
based on the direct construction of the EM self-potentials produced by a
charged spherical particle under the action of an external EM force. In
particular we intend to analyze basic features of the LAD
(Lorentz-Abraham-Dirac) and the LL (Landau-Lifschitz) equations. Both are
shown to lead to incorrect or incomplete results.
\end{abstract}

\date{\today }
\maketitle



\section{Introduction}

An ubiquitous phenomenon which characterizes the dynamics of classical
charged particles is the occurrence of self-forces,\ in particular the
electromagnetic (EM) one which is produced by the EM fields generated by the
same particles. It is well-known that such a force acts on a charged
particle when it is subject also to the action of an arbitrary external
force (see Lorentz, 1892 \cite{Lorentz}; see also for example Landau and
Lifschitz, 1957 \cite{LL}). This phenomenon is usually called as \emph{%
radiation reaction} (\emph{RR}) (Pauli \cite{Pauli1958}) or \emph{radiation
damping} \ (see \cite{Feynman1988}), although a proper distinction between
the two terms can actually be made \cite{Rohrlich2000}. In the sequel we
denote as \emph{RR problem} the treatment of the classical mechanics of a
charged particle in the presence of its EM self-field, and \emph{RR equation}
the corresponding equation of motion for such a particle. The RR problem,
apart its almost endless physical applications, is of fundamental importance
from the theoretical viewpoint for the formulation of consistent
relativistic theories, in particular kinetic theory, plasma dynamics and
astrophysics. Despite efforts more than a century-long, the problem of its "%
\textit{exact}" theoretical description remains still elusive. This refers,
in particular, to the construction of a relativistic equation of motion
which results both \textit{non-perturbative}, in the sense that it does not
rely on a perturbative expansion for the electromagnetic field generated by
the charged particle and \textit{non-asymptotic}, i.e., it does not depend
on any infinitesimal parameter (such, for example, as the radius of the
particle, if it is identified with a spherically-symmetric charge
distribution of finite radius, as in the Lorentz approach \cite{Lorentz}).
The purpose of this paper is to point out that such a problem, since the
historical work of Lorentz which first investigated it \cite{Lorentz}, still
remains essentially unsolved to date. To prove the above statement, in this
paper we intend to analyze in detail the derivation and basic features of
the relativistic equations of motion available so-far, with particular
reference to the so-called LAD and to the LL equations, respectively named
after Lorentz, Abraham and Dirac and Landau and Lifschitz \cite{LL}. A side
motivation of this work is also provided by the recent claim (Rohrlich, 2001
\cite{Rohrlich2001}; see also Spohn, 2000 \cite{Spohn}) that the LL equation
should be regarded as the \textit{exact} (i.e., both non-perturbative and
non-asymptotic) \textit{RR equation}.

\section{A re-formulation of the traditional approach to the RR problem}

In the weakly-relativistic treatment of classical mechanics, the traditional
form of the equations of motion for a charged particle subject to its own EM
self-field can be obtained (see for example Ref.\cite{LL}) by a suitable
asymptotic expansion of the EM self-potentials which generate such a force.
It is instructive to discuss here the basic steps of the derivation. For
definiteness, we shall consider here the case of a single point-particle
with distributed charge density (here denoted as \emph{finite-size charge})%
\emph{\ }carrying constant mass and charge,\emph{\ }$m$ and $q$ (this choice
is actually analogous to that adopted in the original Lorentz approach \cite%
{Lorentz})\emph{.} More precisely, \ it is assumed that the particle mass is
concentrated in the center of the sphere (so that the particle degree of
freedom is the same as that of a point particle), while its charge is
assumed to be uniformly distributed on a spherical shell of finite radius $%
\sigma >0,$ which carries the homogeneous surface charge density $\rho
=q/4\pi \sigma ^{2}$ (\emph{finite-size spherical-shell charge}). Thus,
letting $R=\left\vert \mathbf{r-r}^{\prime }\right\vert$ and denoting
respectively $\left[ \rho (\mathbf{r}^{\prime },t),\mathbf{J}(\mathbf{r}%
^{\prime },t)\right] $ and\ $\left[ \rho (\mathbf{r}^{\prime },t-\frac{R}{c}%
),\mathbf{J}(\mathbf{r}^{\prime },t-\frac{R}{c})\right] $ the charge and
current densities of the particle at time $t$ and at the retarded time $%
t^{\prime }=t-\frac{R}{c},$ we shall impose that
\begin{eqnarray}
\rho (\mathbf{r},t) &=&\frac{q}{4\pi \sigma ^{2}}\delta (\left\vert \mathbf{r%
}-\mathbf{r}(t)\right\vert -\sigma ),  \label{charge densitry} \\
\mathbf{J}(\mathbf{r},t) &=&\frac{q}{4\pi \sigma ^{2}}\overset{\cdot }{%
\mathbf{r}}(t)\delta (\left\vert \mathbf{r}-\mathbf{r}(t)\right\vert -\sigma
).  \label{current
density}
\end{eqnarray}%
Then, if $\mathbf{r}(t)$ is at time $t$ the position of the particle center
of mass,\ the \emph{EM self-force} acting on the same particle is
represented by the Lorentz force acting on the particle itself, i.e.,
\begin{equation}
\mathbf{F}^{(self)}(\mathbf{r}(t),\overset{\cdot }{\mathbf{r}}(t),t)\mathbf{=%
}\int d^{3}\mathbf{r}^{\prime }\left[ \rho (\mathbf{r}^{\prime },t)\mathbf{E}%
^{(self)}(\mathbf{r}^{\prime },t)+\frac{1}{c}\mathbf{J}(\mathbf{r}^{\prime
},t)\times \mathbf{B}^{(self)}(\mathbf{r}^{\prime },t)\right] ,
\end{equation}%
where $\mathbf{F}^{(self)}$ is usually denoted as \emph{RR reaction force}
acting on the particle. Here\ $\left\{ \mathbf{E}^{(self)},\mathbf{B}%
^{(self)}\right\} $ denotes the EM self-field generated by the particle
itself, related to corresponding EM potentials $\left\{ \phi ^{(self)},%
\mathbf{A}^{(self)}\right\} ,$ here denoted as \emph{EM self-potentials}. By
definition $\left\{ \phi ^{(self)},\mathbf{A}^{(self)}\right\} $ \ - to be
evaluated at time $t$ and at the position $\mathbf{r}$ [to be later
identified with the particle center of mass $\mathbf{r}(t)$] - can be
defined as $\phi ^{(self)}(\mathbf{r},t)=\int d^{3}r^{\prime }\frac{1}{R}%
\left[ \rho (\mathbf{r}^{\prime },t-\frac{R}{c})-\rho (\mathbf{r}^{\prime
},t)\right] $ and $\mathbf{A}^{(self)}(\mathbf{r},t)=\frac{1}{c}\int
d^{3}r^{\prime }\frac{1}{R}\left[ \mathbf{J}(\mathbf{r}^{\prime },t-\frac{R}{%
c})-\mathbf{J}(\mathbf{r}^{\prime },t)\right] .$ We stress that the second
terms on the r.h.s. of these equations take into account the charge and
current densities evaluated at the same time $t$ (current time), which
manifestly do not contribute to the self potentials when they are evaluated
at the particle position. In order to evaluate explicitly the EM
self-potentials in this case [Eqs.(\ref{charge densitry}), (\ref{current
density})],\ let us now introduce for them an asymptotic expansion in power
series of $\beta =v/c,$ assuming that $v/c\ll 1.$ Retaining only terms up to
third order (in $\beta $) in $\phi ^{(self)}(\mathbf{r},t)$ and respectively
first order (in $\beta $) in $\mathbf{A}^{(self)}(\mathbf{r},t),$ there
results%
\begin{eqnarray}
\phi ^{(self)}(\mathbf{r},t) &\cong &-\frac{1}{c}\frac{\partial }{\partial t}%
\int d^{3}r^{\prime }\rho (\mathbf{r}^{\prime },t)+\frac{1}{2c^{2}}\frac{%
\partial ^{2}}{\partial t^{2}}\int d^{3}r^{\prime }R\rho (\mathbf{r}^{\prime
},t)-\frac{1}{6c^{3}}\frac{\partial ^{3}}{\partial t^{3}}\int d^{3}r^{\prime
}R^{2}\rho (\mathbf{r}^{\prime },t), \\
\mathbf{A}^{(self)}(\mathbf{r},t) &\cong &-\frac{1}{c}\frac{\partial }{%
\partial t}\int d^{3}r^{\prime }\mathbf{J}(\mathbf{r}^{\prime },t).
\end{eqnarray}%
Thus, introducing the EM gauge $\phi ^{\prime (self)}=\phi ^{(self)}+\frac{1%
}{c}\frac{\partial }{\partial t}f=0,$ $\mathbf{A}^{\prime (self)}(\mathbf{r}%
,t)=\mathbf{A}^{(self)}(\mathbf{r},t)-\nabla f,$ where $f$ is the EM gauge
function (to be denoted for later reference as \emph{RR gauge})
\begin{equation}
f=\int d^{3}r^{\prime }\rho (\mathbf{r}^{\prime },t)-\frac{1}{2c}\frac{%
\partial }{\partial t}\int d^{3}r^{\prime }R\rho (\mathbf{r}^{\prime },t)+%
\frac{1}{6c^{2}}\frac{\partial ^{2}}{\partial t^{2}}\int d^{3}r^{\prime
}R^{2}\rho (\mathbf{r}^{\prime },t),  \label{RR-gauge-3}
\end{equation}%
the transformed vector self-potential $\mathbf{A}^{\prime (self)}(\mathbf{r}%
,t)$ becomes
\begin{equation}
\mathbf{A}^{\prime (self)}(\mathbf{r},t)\cong -\frac{1}{c}\frac{\partial }{%
\partial t}\int d^{3}r^{\prime }\mathbf{J}(\mathbf{r}^{\prime },t)+\nabla
\int d^{3}r^{\prime }\rho (\mathbf{r}^{\prime },t)-\frac{1}{2c}\frac{%
\partial }{\partial t}\nabla ^{\prime }\int d^{3}r^{\prime }R\rho (\mathbf{r}%
^{\prime },t)+\frac{1}{6c^{2}}\frac{\partial ^{2}}{\partial t^{2}}\nabla
^{\prime }\int d^{3}r^{\prime }R^{2}\rho (\mathbf{r}^{\prime },t).
\end{equation}%
After straightforward calculations one obtains%
\begin{equation}
\mathbf{F}^{(self)}(\mathbf{r}(t),\overset{\cdot }{\mathbf{r}}(t),t)=-\frac{%
2q}{3c^{3}}\overset{\cdot \cdot \cdot }{\mathbf{r}}(t)+\alpha q^{2}\left[
\frac{\overset{\cdot \cdot }{\mathbf{r}}(t)}{c^{2}\sigma }+\frac{1}{2\sigma
^{3}}\mathbf{r}(t)\beta ^{2}\right] ,
\end{equation}%
where $\beta ^{2}=\overset{\cdot }{\mathbf{r}^{2}}(t)/c^{2}$ and in the case
of the finite-size spherical shell here considered there results $\alpha =1$%
. Thus, one recovers in this way the \emph{weakly-relativistic RR equation}
or the \emph{weakly-relativistic LAD equation} (after Lorentz \cite{LL},
Abraham \cite{Abraham1905} and Dirac \cite{Dirac1938}) of the form
\begin{equation}
m_{R}\overset{\cdot \cdot }{\mathbf{r}}=\mathbf{F+g}.  \label{a}
\end{equation}%
Here the notation is standard. Thus,
\begin{equation}
\mathbf{g}\equiv -\frac{\mathbf{2}q^{2}}{3c^{3}}\overset{\cdot \cdot \cdot }{%
\mathbf{r}}+\frac{\alpha q^{2}}{2\sigma ^{3}}\mathbf{r}(t)\beta ^{2}
\label{NR self-force}
\end{equation}%
is the \emph{weakly-relativistic \ EM self-force}. Moreover
\begin{equation}
\left\{
\begin{array}{c}
\mathbf{F=q}\left[ \mathbf{E}^{(ext)}+\frac{1}{c}\overset{\cdot }{\mathbf{r}}%
\times \mathbf{B}^{(ext)}\right] , \\
m_{R}\equiv m-m_{EM}, \\
m_{EM}\equiv \alpha \frac{q^{2}}{c^{2}\sigma },%
\end{array}%
\right.  \label{Lorentz force}
\end{equation}%
are respectively the Lorentz force produced by the external EM field, the
\emph{renormalized mass }and the so-called \emph{EM mass. Eq.(\ref{a})
manifestly indicates that the limit }$\sigma \rightarrow 0$\emph{\
(point-particle) does not exist. Hence the LAD equation is only defined for
a finite-size spherical-shell charge, i.e., for which it results }$\sigma >0$
\emph{and possibly also} $m_{R}>0$ (the vanishing of $m_{R}$ involves in
fact the manifest violation of Newton's second law)\emph{.} The derivation
of the analogous \emph{relativistic RR equation} is usually achieved (see
for example the treatment given in Ref. \cite{LL}) under the implicit
assumptions that: \emph{1) higher order corrections in }$\beta $\emph{\ are
negligible; 2) all previous asymptotic expansions in power series of }$\beta
$\emph{\ remain valid for the whole range of values of particle velocities,
i.e., even if }$\beta \sim o(1)$\emph{, namely arbitrary close to the speed
of light. }In practice, the construction of the equation is achieved
formally by writing Eq.(\ref{a}) in covariant form and by suitably replacing
the definitions of the renormalized mass $m_{R}$ and of the self-force $%
\mathbf{g}$ by an appropriate Lorentz scalar $m_{oR}$ and a 4-vector $g^{\mu
}.$ Both are to be defined in such a way that in the weakly-relativistic
limit they recover the correct values set by the previous equations. The
relativistic RR equation of motion thus obtained, nowadays popularly known
as the \emph{relativistic LAD equation,}

\begin{equation}
m_{oR}c\frac{d^{2}r^{\mu }}{ds^{2}}=\frac{q}{c}F^{\mu \nu }u^{\nu }+g^{\mu }
\label{relativistic LAD}
\end{equation}%
was first presented by Dirac in his famous paper on relativistic radiation
reaction in classical electrodynamics \cite{Dirac1938}. Here, $g^{\mu }$ is
the EM self-force which reads%
\begin{equation}
g^{\mu }=\frac{2q^{2}}{3c}\left\{ \frac{d^{2}u^{\mu }}{ds^{2}}-u^{\mu
}u^{\nu }\frac{d^{2}u_{\nu }}{ds^{2}}\right\} .  \label{LAD g_mu}
\end{equation}%
Furthermore\ the relativistic renormalized mass has the form
\begin{equation}
m_{oR}=m_{o}+m_{EM}+\Delta m_{EM},  \label{renormalized rest mass}
\end{equation}%
with $m_{o},$ $m_{EM}$ and $\Delta m_{EM}$ respectively , the rest mass, the
EM mass and a suitable relativistic correction. Finally, as usual $r^{\mu },$
$u^{\mu }$ and $F^{\mu \nu }$ denote the 4-position vector, the 4-velocity
vector and the Faraday tensor associated to the external EM field $%
A^{(ext)\mu }.$ \ Nevertheless, \textit{the LAD equation is valid only in
the case of the Minkowsky metric}. Its generalization for arbitrary curved
space-time was later carried out by DeWitt and Brehme \cite{DeWitt}.

\section{Difficulties with RR equations}

Since Lorentz famous paper \cite{Lorentz} many textbooks and research
articles have appeared on the subject of RR. Among them are \cite%
{Rohrlich1965,Teitelboim1970,Sokolov1986,Parrott1987,Parrott1993}, where one
can find the discussion of the related problems: mass renormalization,
non-uniqueness, runaway solutions and pre-acceleration. In the literature
the usual derivations of relativistic form of the EM self-force are made
under the implicit\ assumption that all the expansions in powers used near
the particle trajectory are valid for the whole range of values of particle
velocity, in particular, arbitrary close to that of the light. But it is
easy to see that this is not true in general.

It is often said that the LAD equation, in its weakly-relativistic form
given by Eq.(\ref{a}), is unsatisfactory because it requires the
specification of the initial acceleration $\overset{\cdot \cdot }{\mathbf{r}}%
(t_{o}),$ besides the initial state $\left\{ \mathbf{r}(t_{o}),\overset{%
\cdot }{\mathbf{r}}(t_{o})\right\} .$ As a consequence it violates the
Newton's principle of determinacy (NPD), one of the building blocks of
classical mechanics. Moreover, another critical issue arises due to the
appearance of so-called \emph{runaway solutions}. \ These are solutions
which by definitions blow up in time for $t\rightarrow +\infty .$ One can
show that they are present for arbitrary external forces $\mathbf{F.}$ In
particular,\ in the case $\mathbf{F}\equiv \mathbf{0}$ the general solution
of Eq.(\ref{a}) takes the form $\mathbf{r}(t)\mathbf{=}\overset{\cdot \cdot }%
{\mathbf{r}}(t_{o})\exp \left\{ \frac{(t-t_{o})}{\tau }\right\} ,$where $%
\tau >0$ is the constant parameter $\tau =\frac{\mathbf{2q}^{2}}{3mc^{3}}.$
This equation is clearly unsatisfactory. In fact, it violates also another
fundamental principle of classical mechanics, the Galilei law of inertia,
according to which an isolated particle must have a constant velocity in any
inertial Galilean frame. In an attempt to circumvent this difficulty it has
been suggested to replace Eq.(\ref{a}) by a suitably-modified integral (or
differential) equation. An example is provided by an integro-differential
equation which in its weakly-relativistic form reads \cite%
{Haag1955,Rohrlich1965}%
\begin{equation}
m_{R}\overset{\cdot \cdot }{\mathbf{r}}=\int\limits_{0}^{\infty }\mathbf{F}%
(t+\tau s)e^{-s}ds\equiv \frac{1}{\tau }\int\limits_{t}^{\infty }\mathbf{F}%
(x)e^{-\frac{x-t}{\tau }}dx  \label{c}
\end{equation}%
(\emph{Haag equation}). This equation is manifestly consistent with PND and
the law of inertia (since runaway solutions are excluded). However, it
exhibits another serious drawback: it violates the principle of causality.
Indeed, contrary to the requirement according to which the effect cannot
precede the cause, the paradox of \emph{pre-acceleration} is implied by Eq.(%
\ref{c}). Accordingly, a particle obeying Eq.(\ref{c}) would experience a
force \emph{before} actually turning it on! \ Eq.(\ref{c}) clearly violates
this principle since if a constant force $\mathbf{F}_{o}$ is suddenly turned
on at $t_{o}$ letting $\mathbf{F}(x)=0$ for $x<t_{o}$ and $\mathbf{F}(x)=%
\mathbf{F}_{o}$ for $x\geq t_{o}$ (\emph{sudden force}) it follows%
\begin{equation}
\left\{
\begin{array}{ccc}
\frac{1}{\tau }\int\limits_{t}^{\infty }\mathbf{F}(x)e^{-\frac{x-t}{\tau }%
}dx=\mathbf{F}_{o}\frac{1}{\tau }\int\limits_{t_{o}}^{\infty }e^{-\frac{x-t}{%
\tau }}dx=\mathbf{F}_{o}e^{-\frac{t_{o}-t}{\tau }} &  & t<t_{o}, \\
\frac{1}{\tau }\int\limits_{t}^{\infty }\mathbf{F}(x)e^{-\frac{x-t}{\tau }%
}dx=\mathbf{F}_{o}\frac{1}{\tau }\int\limits_{t}^{\infty }e^{-\frac{x-t}{%
\tau }}dx=\mathbf{F}_{o} &  & t\geq t_{o}.%
\end{array}%
\right.
\end{equation}%
Hence, the RR equation before and after $t_{o}$ will be respectively
provided by%
\begin{equation}
m_{R}\overset{\cdot \cdot }{\mathbf{r}}=\left\{
\begin{array}{ccc}
\mathbf{F}_{o}e^{-\frac{t_{o}-t}{\tau }} &  & t<t_{o}, \\
\mathbf{F}_{o} &  & t\geq t_{o}.%
\end{array}%
\right.
\end{equation}%
Hence, the particle should actually experience an acceleration for all $%
t<t_{o}$, a phenomenon usually denoted as pre-acceleration. For this reason
neither Eq.(\ref{a}) nor (\ref{c}) can be considered satisfactory RR
equations.

Another attempt is to adopt an "iterative approach" whereby the time
derivative of the acceleration appearing in the self-force (\ref{NR
self-force}) is approximated in terms of the time derivative of the external
force (this approach is manifestly admissible only if the self-force can be
treated as suitably small). In the weakly-relativistic LAD, letting first $%
\frac{2q}{3c^{3}}\overset{\cdot \cdot \cdot }{\mathbf{r}}(t)\overset{\cdot
\cdot }{\mathbf{r}}\cong \tau \frac{d}{dt}\mathbf{F,}$ one obtains the
reduced equation $m_{R}\overset{\cdot \cdot }{\mathbf{r}}=\mathbf{F+}\tau
\frac{d}{dt}\mathbf{F.}$ More generally, by carrying out the full iteration
in Eq.(\ref{a}), one obtains the equation $m_{R}\overset{\cdot \cdot }{%
\mathbf{r}}=\mathbf{F+}\sum\limits_{n=1}^{\infty }\tau ^{n}\frac{d^{n}}{%
dt^{n}}\mathbf{F.}$ This equation, and the analogous relativistic one which
follows in a similar way from Eq.(\ref{relativistic LAD}), are however
useless since they are infinite-order ode's (and hence have no solutions!).
\

Nevertheless, this difficult can be circumvented by introducing a "reduction
method" whereby all the derivatives higher than $\overset{\cdot }{\mathbf{r}}
$ appearing in each term $\frac{d^{n}}{dt^{n}}\mathbf{F}$ are expressed via
the same iterative approach which permits to cast $\frac{d^{n}}{dt^{n}}%
\mathbf{F}$, for each $n\geq 1$ in the form $\frac{d^{n}}{dt^{n}}\mathbf{F=F}%
_{n}(\mathbf{r},\overset{\cdot }{\mathbf{r}},t).$ This idea has been
exploited by Cook \cite{Cook1983} who obtained the so-called \emph{Cook RR
equation}%
\begin{equation}
m_{R}\overset{\cdot \cdot }{\mathbf{r}}=\mathbf{F+}\sum\limits_{n=1}^{\infty
}\tau ^{n}\mathbf{F}_{n}(\mathbf{r},\overset{\cdot }{\mathbf{r}},t)\mathbf{.}
\end{equation}%
\bigskip This procedure was actually first adopted by Landau and Lifschitz
\cite{LL} based on a one-step iteration of the form
\begin{equation}
m_{R}\overset{\cdot \cdot }{\mathbf{r}}\cong \mathbf{F+}\tau \mathbf{F}_{1}(%
\mathbf{r},\overset{\cdot }{\mathbf{r}},t)
\end{equation}%
(which could be denoted as the \emph{weakly-relativistic} \emph{LL equation }%
\cite{LL}). In the corresponding relativistic formulation (see Ref. \cite{LL}%
), applying the one-step reduction process to the relativistic LAD equation (%
\ref{relativistic LAD}) delivers for $g^{\mu }$ [as given by Eq.(\ref{LAD
g_mu})] the approximation
\begin{eqnarray}
g^{\mu } &\cong &\frac{2q^{3}}{3c^{3}}\frac{\partial F^{\mu k}}{\partial
r^{i}}u_{k}u^{i}-\frac{2q^{4}}{3m_{R}c^{5}}F^{\mu k}F_{jk}u^{j}+
\label{reduced self-force} \\
&&+\frac{2q^{4}}{2m_{R}^{2}c^{5}}F_{kl}u^{l}F^{km}u_{m}u^{\mu }.  \notag
\end{eqnarray}

The relativistic LL equation obtained in such a way by Landau and Lifschitz
\emph{\ }\cite{LL} involves, however, also the replacement of the
renormalized mass $m_{oR}$ with the inertial mass only $m_{o}$. While in the
framework on classical electrodynamics the latter position remains totally
unjustified, the resulting equation has been claimed by Rohrlich \cite%
{Rohrlich2001}) to be the \emph{exact RR equation. }Also in view of the
previous discussion, the validity of this statement seems unlikely. In fact
the LL equation (both in the weakly-relativistic and fully relativistic
versions) has remaining serious problems:

\begin{itemize}
\item One reason is that as recalled in Sec.2 the theory is asymptotic in $%
\beta $ and does not take into account in a consistent way higher-order
finite-$\beta $ effects;

\item A second motivation is that it only applies provided the external
force is suitably smooth, i.e., of class $C^{2}$. In fact, if an external
force is turned on suddenly (\emph{sudden force}), as in the example
provided above for the weakly-relativistic case, the LL equation becomes
manifestly invalid. Since sudden forces cannot be ruled out, purely on first
principles, this is actually a major conceptual difficulty (and potential
inconsistency) of the part of the LL equation.

\item An additional difficulty of the LL\ equation is that it is \emph{%
non-variational}, i.e., the Hamilton variational principle does not apply
for such an equation, even if the external force is identified with the
Lorentz force [see Eq.(\ref{Lorentz force})]. \ This feature is actually
common to all current RR equations. This is illustrated, for example, by the
weakly-relativistic LAD equation Eq.(\ref{a}) which appears manifestly
non-variational. In fact in this case the Lagrangian function $L$ should
actually depend not only on the Lagrangian state $(\mathbf{r,}\overset{\cdot
}{\mathbf{r}})$\textbf{\ }but also on the acceleration\textbf{\ }$\overset{%
\cdot \cdot }{\mathbf{r}}$. Thus, for example, letting introducing the
Lagrangian $L(\mathbf{r,}\overset{\cdot }{\mathbf{r}},\overset{\cdot \cdot }{%
\mathbf{r}},t)=\frac{1}{2}m_{R}\overset{\cdot }{\mathbf{r}}^{2}-q\left[ \phi
^{(ext)}(\mathbf{r,}t)-\frac{1}{c}\overset{\cdot }{\mathbf{r}}\cdot \mathbf{A%
}^{(ext)}(\mathbf{r,}t)\right] +f(\mathbf{r,}\overset{\cdot }{\mathbf{r}},%
\overset{\cdot \cdot }{\mathbf{r}},t),$ for the validity of the Hamilton
variational principle it should be%
\begin{eqnarray}
&&\left. \delta \int\limits_{t_{1}}^{t_{2}}dtL(\mathbf{r}(t)\mathbf{,}%
\overset{\cdot }{\mathbf{r}}(t),\overset{\cdot \cdot }{\mathbf{r}}%
(t),t)=\int\limits_{t_{1}}^{t_{2}}dt\delta \mathbf{r}(t)\cdot \left[ m_{R}%
\overset{\cdot \cdot }{\mathbf{r}}-\mathbf{F}\right] +\right. \\
&&\left. +\int\limits_{t_{1}}^{t_{2}}dt\delta f(\mathbf{r}(t)\mathbf{,}%
\overset{\cdot }{\mathbf{r}}(t),\overset{\cdot \cdot }{\mathbf{r}}%
(t),t)=0,\right.  \notag
\end{eqnarray}%
where $\delta f(\mathbf{r}(t)\mathbf{,}\overset{\cdot }{\mathbf{r}}(t),%
\overset{\cdot \cdot }{\mathbf{r}}(t),t)$ is an exact differential form such
that $\int\limits_{t_{1}}^{t_{2}}dt\delta f(\mathbf{r}(t)\mathbf{,}\overset{%
\cdot }{\mathbf{r}}(t),\overset{\cdot \cdot }{\mathbf{r}}(t),t)=\int%
\limits_{t_{1}}^{t_{2}}dt\delta \mathbf{r}(t)\cdot \frac{2q^{2}}{3c^{3}}%
\overset{\cdot \cdot \cdot }{\mathbf{r}}(t).$ This implies, however,
\begin{equation}
\int\limits_{t_{1}}^{t_{2}}dt\delta f(\mathbf{r}(t)\mathbf{,}\overset{\cdot }%
{\mathbf{r}}(t),\overset{\cdot \cdot }{\mathbf{r}}(t),t)=-\frac{q^{2}}{3c^{3}%
}\int\limits_{t_{1}}^{t_{2}}dt\left[ \delta \overset{\cdot }{\mathbf{r}}%
(t)\cdot \overset{\cdot \cdot }{\mathbf{r}}(t)-\delta \overset{\cdot \cdot }{%
\mathbf{r}}(t)\cdot \overset{\cdot }{\mathbf{r}}(t)\right] \equiv 0,
\end{equation}%
which means that a real function $f(\mathbf{r}(t)\mathbf{,}\overset{\cdot }{%
\mathbf{r}}(t),\overset{\cdot \cdot }{\mathbf{r}}(t),t)$ fulfilling the
previous constraint cannot exist.

\item Finally, another serious difficulty is related also to the conditions
validity of the reduction process indicated above. In fact, it is obvious
that the one-step reduction adopted for the derivation of the LL equation
provides, at most, an asymptotic approximation for the (still unknown) exact
RR equation.
\end{itemize}

\section{Conclusions}

In this paper we have reviewed aspects of the RR problem. We have shown that:

\begin{itemize}
\item the difficulties with the LAD equation are intrinsic, i.e., arise due
to the fact that the equation is a third order ode. As a consequence, the
Newton's principle of determinacy and the Galilei law of inertia are
potentially violated;

\item attempts to circumvent these difficulties, based on various approaches
(Haag,\emph{\ }LL and Cook equations) fail for different reasons;

\item limitations of the current RR equations have been pointed out, with
particular reference to the conditions of validity of the LL equation;

\item all current RR equations are non-variational, implying - contrary to
common knowledge in classical mechanics - that the dynamics of a charged
particle described by these model equations does not define an Hamiltonian
system. This is the reason why relativistic systems of charged particle
usually are not (or cannot be) described as Hamiltonian systems.
Nevertheless, it not clear whether this feature is only an accident, i.e.,
is only due to the approximations introduced in the RR equation or is
actually intrinsic to the nature of the RR problem.  Unfortunately, the
precise solution of this dilemma is not yet known, although the prevailing
opinion seems directed to the first possibility (see for example Ref. \cite%
{Kosyakov2007}). For a proper discussion of the issue we refer to the
accompanying paper \cite{Dorigo2008a}.
\end{itemize}

This suggests, in our view, that in many respects the RR problem is still
open. Its solution should be based on the search of a relativistic, \textit{%
non-perturbative equation of motion for a particle in the presence of its EM
self-field}. This problem poses not just an intellectual challenge but a
fundamental physical requirement in all applications which involve the
description of relativistic dynamics of classical charged particles, such as
relativistic kinetic theory, plasma physics and astrophysics. In an
accompanying paper an attempt at a possible solution of the\ RR problem will
be discussed \cite{Dorigo2008a}.

\section*{Acknowledgments}

Work developed in cooperation with the CMFD Team, Consortium for
Magneto-fluid-dynamics (Trieste University, Trieste, Italy). \ Research
developed in the framework of the MIUR (Italian Ministry of University and
Research) PRIN Programme: \textit{Modelli della teoria cinetica matematica
nello studio dei sistemi complessi nelle scienze applicate}. The support
(A.B) of ICTP (International Center for Theoretical Physics, Trieste,
Italy), (A.B.) University of Trieste, Italy, (M.T) COST Action P17 (EPM,
\textit{Electromagnetic Processing of Materials}) and (M.T. and P.N.) GNFM
(National Group of Mathematical Physics) of INDAM (Italian National
Institute for Advanced Mathematics) is acknowledged.

\section*{Notice}

$^{\S }$ contributed paper at RGD26 (Kyoto, Japan, July 2008). \newpage



\end{document}